\documentclass[manuscript]{aastex}

\usepackage{hyperref}
\usepackage{url}
\usepackage{lscape}
\usepackage{longtable}
\usepackage{latexsym}
\usepackage{graphics}
\usepackage{graphicx}
\usepackage{enumitem}
\usepackage{amsmath}
\usepackage{mathptmx}
\usepackage{latexsym}
\usepackage{fix-cm}
\usepackage{bigstrut}
\usepackage{booktabs}

\begin{document}

\title{A New Class of Anisotropic Charged Compact Star}

\author{B. S. Ratanpal\altaffilmark{1}}
\affil{Department of Applied Mathematics, Faculty of Technology \& Engineering, The M. S. University of Baroda, Vadodara - 390 001, India}
\email{bharatratanpal@gmail.com}

\and

\author{Piyali Bhar\altaffilmark{2}}
\affil{Department of Mathematics, Government General Degree College, Singur, Hooghly, West
Bengal-712409, India}
\email{piyalibhar90@gmail.com}

\begin{abstract}

A new model of charged compact star is reported by solving the Einstein-Maxwell field equations by choosing a suitable form of 
radial pressure. The model parameters $\rho$, $p_r$, $p_{\perp}$ and 
$E^{2}$  are in closed form and all are well behaved inside the stellar interior. A comparative study of charged and 
uncharged model is done with the help of graphical analysis.

\end{abstract}


\section{Introduction}
\label{sec:1}
To find the exact solution of Einstein's field equations is difficult due to its non-linear nature. A large number of exact 
solutions of Einstein’s field equations in literature but not all of them are physically
relevant. A comprehensive collection of static, spherically symmetric solutions are found in \cite{step} and \cite{lake}.
%%%%%%%%%%%%%%%%%%%%%%% charged %%%%%%%%%%%%%%%%%%%%%%%%%%%%
A large collection of models of stellar objects incorporating charge can be found in literature.
\cite{stet} proposed that a fluid sphere of uniform density with a net surface charge is more stable than without charge. An
interesting observation of \cite{kras} is that in the presence of charge, the gravitational collapse of a spherically symmetric 
distribution of matter to a point singularity may be avoided. Charged anisotropic matter with
linear equation of state is discussed by \cite{maharaj}. \cite{joshi} found that the solutions of Einstein-Maxwell system of 
equations are important to study the cosmic censorship hypothesis and
the formation of naked singularities. The presence of charge affects
the values for redshifts, luminosities, and maximum mass for stars. Charged perfect fluid sphere satisfying a linear
equation of state was discussed by \cite{ivanov}. Regular models with quadratic equation of state was discussed by \cite{takisa}.
They obtained exact and physically reasonable solution of Einstein-Maxwell system of equations. Their model is well behaved and
regular. In particular there is no singularity in the proper charge density. \cite{varela} considered a self gravitating, charged and anisotropic fluid
sphere. To solve Einstein-Maxwell field equation they have assumed both linear and nonlinear equation of state and
discussed the result analytically. \cite{feroze} extend the work of \cite{thiru} by considering quadratic equation of state for 
the matter distribution to study the general situation of a compact relativistic body in presence of electromagnetic field and anisotropy.

%%%%%%%%%%%%%%%%%%%%%%%%%%%%%%%%%%%%%%%%%  anisotropy %%%%%%%%%%%%%%%%%%%%%%%%%%%%%%
\noindent \cite{rud72} investigated that for highly compact
astrophysical objects like X-ray pulsar, Her-X-1, X-ray
buster 4U 1820-30, millisecond pulsar SAX J 1804.4-3658, PSR J1614-2230, LMC X-4
etc. having core density beyond the nuclear density
$(\sim~10^{15}gm/cm^{3})$ there can be pressure anisotropy, i.e, the pressure inside
these compact objects can be decomposed into
two parts radial pressure $p_r$ and transverse pressure
$p_\perp$ perpendicular direction to $p_r$.
$\Delta=p_r-p_\perp$ is called the anisotropic factor which measures the anisotropy. The reason
behind these anisotropic nature are the existence of
solid core, in presence of type 3A superfluid \cite{kip}, phase transition \cite{sokolov}, pion condensation \cite{saw},
rotation, magnetic field, mixture of two fluid,
existence of external field etc. Local anisotropy in self gravitating
systems were studied by \cite{herrera97}. \cite{dev} demonstrated that
pressure anisotropy affects the physical properties, stability
and structure of stellar matter.
Relativistic stellar model admitting a quadratic equation
of state was proposed by \cite{sharma13} in finch-skea spacetime. \cite{panda14} has generalized earlier work in
modified Finch-Skea spacetime by
incorporating a dimensionless parameter n. In a
very recent work \cite{piy15a} obtained a new model of
an anisotropic superdense star which admits conformal
motions in the presence of a quintessence field which is
characterized by a parameter $\omega_q$ with $-1 < \omega_q < -1/3$.
The model has been developed by choosing \cite{vaidya82} {\em ansatz}. \cite{murad} have studied the behavior of
static spherically symmetric relativistic objects with locally
anisotropic matter distribution considering the Tolman VII
form for the gravitational potential $g_{rr}$ in curvature coordinates
together with the linear relation between the energy
density and the radial pressure.\\
\noindent Charged anisotropic star on paraboloidal spacetime was studied by \cite{ratanpal1}. \cite{ratanpal2} studied anisotropic
star on pseudo-spheroidal spacetime. Charged anisotropic star on pseudo-spheroidal spacetime was studied by \cite{ratanpal3}. The study
of compact stars having Matese and Whitman mass function was carried out by \cite{ratanpal4}.
Motivated by these earlier works in the present paper we develop a model of compact star by incorporating charge. 
Our paper is organized as follows: In section 2, interior spacetime and the Einstein-Maxwell system is discussed. Section 3 deals 
with solution of field equations. Section 4 contains exterior spacetime and matching conditions. Physical analysis of the model is
discussed in section 5. Section 6 contains conclusion.
\section{Interior Spacetime}
\label{sec:2}
\noindent We consider the static spherically symmetric spacetime metric as,
\begin{equation}\label{IMetric1}
	ds^{2}=e^{\nu(r)}dt^{2}-e^{\lambda(r)}dr^{2}-r^{2}\left(d\theta^{2}+\sin^{2}\theta d\phi^{2} \right).
\end{equation}
Where $\nu$ and $\lambda$ are functions of the radial coordinate `r' only.\\
Einstein-Maxwell Field Equations is given by
\begin{equation}\label{FE}
	R_{i}^{j}-\frac{1}{2}R\delta_{i}^{j}=8\pi\left(T_{i}^{j}+\pi_{i}^{j}+E_{i}^{j} \right),
\end{equation}
where,
\begin{equation}\label{Tij}
	T_{i}^{j}=\left(\rho+p \right)u_{i}u^{j}-p\delta_{i}^{j},
\end{equation}
\begin{equation}\label{piij}
	\pi_{i}^{j}=\sqrt{3}S\left[c_{i}c^{j}-\frac{1}{2}\left(u_{i}u^{j}-\delta_{i}^{j} \right) \right],
\end{equation}
and
\begin{equation}\label{Eij}
	E_{i}^{j}=\frac{1}{4\pi}\left(-F_{ik}F^{jk}+\frac{1}{4}F_{mn}F^{mn}\delta_{i}^{j} \right).
\end{equation}
Here $\rho$ is proper density, $p$ is fluid pressure, $u_{i}$ is unit four velocity, $S$ denotes magnitude of anisotropic tensor
and $C^{i}$ is radial vector given by $\left(0,-e^{-\lambda/2},0,0 \right)$. $F_{ij}$ denotes the anti-symmetric
electromagnetic field strength tensor defined by
\begin{equation}\label{Fij}
	F_{ij}=\frac{\partial A_{j}}{\partial x_{i}}-\frac{\partial A_{i}}{\partial x_{j}},
\end{equation}
that satisfies the Maxwell equations
\begin{equation}\label{ME1}
	F_{ij,k}+F_{jk,i}+F_{ki,j}=0,
\end{equation}
and
\begin{equation}\label{ME2}
	\frac{\partial}{\partial x^{k}}\left(F^{ik}\sqrt{-g} \right)=4\pi\sqrt{-g}J^{i},
\end{equation}
where $g$ denotes the determinant of $g_{ij}$, $A_{i}=\left(\phi(r), 0, 0, 0 \right)$ is four-potential and
\begin{equation}\label{Ji}
	J^{i}=\sigma u^{i},
\end{equation}
is the four-current vector where $\sigma$ denotes the charge density.

The only non-vanishing components of $F_{ij}$ is $F_{01}=-F_{10}$. Here
\begin{equation}\label{F01}
	F_{01}=-\frac{e^{\frac{\nu+\lambda}{2}}}{r^{2}}\int_{0}^{r} 4\pi r^{2}\sigma e^{\lambda/2}dr,
\end{equation}
and the total charge inside a radius $r$ is given by
\begin{equation}\label{qr}
	q(r)=4\pi\int_{0}^{r} \sigma r^{2}e^{\lambda/2}dr.
\end{equation}
The electric field intensity $E$ can be obtained from $E^{2}=-F_{01}F^{01}$, which subsequently reduces to
\begin{equation}\label{E}
	E=\frac{q(r)}{r^{2}}.
\end{equation}
The field equations given by (\ref{FE}) are now equivalent to the following set of the non-linear ODE's
\begin{equation}\label{FE1}
	\frac{1-e^{-\lambda}}{r^{2}}+\frac{e^{-\lambda}\lambda'}{r}=8\pi\rho+E^{2},	
\end{equation}
\begin{equation}\label{FE2}
	\frac{e^{-\lambda}-1}{r^{2}}+\frac{e^{-\lambda}\nu'}{r}=8\pi p_{r}-E^{2},
\end{equation}
\begin{equation}\label{FE3}
	e^{-\lambda}\left(\frac{\nu''}{2}+\frac{\nu'^{2}}{4}-\frac{\nu'\lambda'}{4}+\frac{\nu'-\lambda'}{2r} \right)=8\pi p_{\perp}+E^{2},
\end{equation}
where we have taken
\begin{equation}\label{pr1}
	p_{r}=p+\frac{2S}{\sqrt{3}},
\end{equation}
\begin{equation}\label{pp1}
	p_{\perp}=p-\frac{S}{\sqrt{3}}.
\end{equation}
\begin{equation}\label{S}
	8\pi\sqrt{3}S=p_{r}-p_{\perp}.
\end{equation}
\section{Solution of Field Equations}
To solve the above set of equations (\ref{FE1})-(\ref{FE3}) we take the mass function of the form
\begin{equation}\label{mass}
m(r)=\frac{br^{3}}{2(1+ar^{2})},
\end{equation}
where `a' and `b' are two positive constants. The mass function given in (\ref{mass}) is known as Matese \& Whitman \cite{mat80}
 mass function that gives a monotonic decreasing matter density which was used by \cite{mak03} to model an anisotropic fluid star,
\cite{lobo06} to develop a model of dark energy star, \cite{sharma07} to model a class of relativistic stars with a linear 
equation of state and \cite{maharaj08} to model a charged anisotropic matter with linear equation of state.\\
Using the relationship $e^{-\lambda}=1-\frac{2m}{r}$ and equation (\ref{mass}) we get,

\begin{equation}\label{elambda}
	e^{\lambda}=\frac{1+ar^{2}}{1+(a-b)r^{2}}.
\end{equation}
From equation (\ref{FE1}) and (\ref{elambda}) we obtain
\begin{equation}\label{rho1}
	8\pi\rho=\frac{3b+abr^{2}}{(1+ar^{2})^{2}}-E^{2}.
\end{equation}
We choose $E^{2}$ of the form
\begin{equation}\label{E2}
	E^{2}=\frac{\alpha ar^{2}}{(1+ar^{2})^2},
\end{equation}
which is regular at the center of the star.
Substituting the expression of $E^{2}$ into (\ref{rho1}) we get,
\begin{equation}\label{rho2}
	8\pi\rho=\frac{3b+a(b-\alpha)r^{2}}{(1+ar^{2})^{2}}.
\end{equation}
To integrate the equation (\ref{FE2}) we take radial pressure of the form,
\begin{equation}\label{pr2}
	8\pi p_{r}=\frac{bp_{0}(1-ar^{2})}{(1+ar^{2})^{2}},
\end{equation}
where $p_{0}$ is a positive constant, the choice of $p_r$ is reasonable due to the fact that it is monotonic decreasing function 
of `r' and the radial pressure vanishes at $r=\frac{1}{\sqrt{a}}$ which gives the radius of the star.\\
From (\ref{pr2}) and (\ref{FE2}) we get,
\begin{equation}\label{nudash}
	\nu'=\frac{(bp_{0}+b)r-a(bp_{0}+\alpha-b)r^{3}}{(1+ar^{2})\left[1+(a-b)r^{2} \right]}.
\end{equation}
Integrating we get,
\begin{equation}\label{nu}
	\nu=log\left\{\frac{C\left(1+ar^{2} \right)^{\left(\frac{2bp_{0}+\alpha}{2b} \right)} }{\left[\left(b-a \right)r^{2}-1 \right]^{\left[\frac{\left(b^{2}-2ab \right)p_{0}+b^{2}-\alpha a }{2b^{2}-2ab} \right]} } \right\},
\end{equation}
where $C$ is contant of integration, and the spacetime metric in the interior is given by
\begin{equation}\label{IMetric2}
	ds^{2}=\left\{\frac{C\left(1+ar^{2} \right)^{\left(\frac{2bp_{0}+\alpha}{2b} \right)} }{\left[\left(b-a \right)r^{2}-1 \right]^{\left[\frac{\left(b^{2}-2ab \right)p_{0}+b^{2}-\alpha a }{2b^{2}-2ab} \right]} } \right\} dt^{2}-\left[\frac{1+ar^{2}}{1+\left(a-b \right)r^{2}} \right]dr^{2}-r^{2}\left(d\theta^{2}+\sin^{2}\theta d\phi^{2} \right).
\end{equation}
\noindent From (\ref{FE2}), (\ref{FE3}) and (\ref{S}), we have
\begin{equation}\label{S1}
	8\pi\sqrt{3}S=\frac{r^{2}\left[A_{1}+A_{2}r^{2}+A_{3}r^{4} \right]}{\left[-4+B_{1}r^{2}+B_{2}r^{4}+B_{3}r^{6}+B_{4}r^{8} \right]},
\end{equation}
where $A_{1}=b^{2}p_{0}^{2}+14b^{2}p_{0}-12abp_{0}+3b^{2}-12\alpha a$,\\
 $A_{2}=-2ab^{2}p_{0}^{2}+8ab^{2}p_{0}-8a^{2}bp_{0}-2\alpha abp_{0}+2ab^{2}+8\alpha ab-16\alpha a^{2}$,\\
$A_{3}=a^{2}b^{2}p_{0}^{2}-4a^{2}b^{2}p_{0}+4a^{3}bp_{0}+2\alpha a^{2}bp_{0}-a^{2}b^{2}+4\alpha a^{2}b-4\alpha a^{3}+\alpha^{2}a^{2}$,\\
$B_{1}=4b-16a$,~~~~~~~ $B_{2}=12ab-24a^{2}$, ~~~~~~~$B_{3}=12a^{2}b-16a^{3}$ and~~~~~~~~ $B_{4}=4a^{3}b-4a^{4}$.

\noindent From (\ref{S}) we obtain,
\begin{equation}\label{pp2}
	8\pi p_{\perp}=\frac{\left[4bp_{0}+C_{1}r^{2}+C_{2}r^{4}+C_{3}r^{6}\right]}{\left[4-B_{1}r^{2}-B_{2}r^{4}-B_{3}r^{6}-B_{4}r^{8} \right]},
\end{equation}
where, $C_{1}=b^{2}p_{0}^{2}-8abp_{0}+3b^{2}-12\alpha a$, \\
$C_{2}=-2ab^{2}p_{0}^{2}+8ab^{2}p_{0}-12a^{2}bp_{0}-2\alpha abp_{0}+2ab^{2}+8\alpha ab-16\alpha a^{2}$,\\
$C_{3}=a^{2}b^{2}p_{0}^{2}+2\alpha a^{2}bp_{0}-a^{2}b^{2}+4\alpha a^{2}b-4\alpha a^{3}+\alpha^{2}a^{2}$.

\section{Exterior Spacetime and Matching Condition}
\label{sec:3}
\noindent we match our interior spacetime (\ref{IMetric2}) to the exterior Reissner-Nordstr\"{o}m spacetime at the boundary $r=r_b$ (where $r_b$ is the radius of the star.). The exterior spacetime is given by the line element
\begin{equation}\label{EMetric}
	ds^{2}=\left(1-\frac{2M}{r}+\frac{q^{2}}{r^{2}} \right)dt^{2}-\left(1-\frac{2M}{r}+\frac{q^{2}}{r^{2}} \right)^{-1}dr^{2}-r^{2}\left(d\theta^{2}+\sin^{2}\theta d\phi^{2} \right).
\end{equation}

\noindent By using the continuity of the metric potential $g_{rr}$ and $g_{tt}$ at the boundary $r=r_b$ we get,
\begin{equation}
e^{\nu(r_b)}=1-\frac{2M}{r_b}+\frac{q^{2}}{r^{2}},
\end{equation}
\begin{equation}
e^{\lambda(r_b)}=\left(1-\frac{2M}{r_b}+\frac{q^{2}}{r^{2}}\right)^{-1}.
\end{equation}
The radial pressure should vanish at the boundary of the star, hence from equation (\ref{pr2}) we obtain
\begin{equation}\label{a}
a=\frac{1}{r_b^{2}}.
\end{equation}
\noindent Using (\ref{a}) \& (\ref{mass}) we obtain
\begin{equation}\label{b}
b=\frac{4m}{r_b^{3}}.
\end{equation}
We compute the values of `a' and `b' for different compact stars which is given in table \ref{table1}.\\
\begin{figure}[htbp]
    \centering
        \includegraphics[scale=.3]{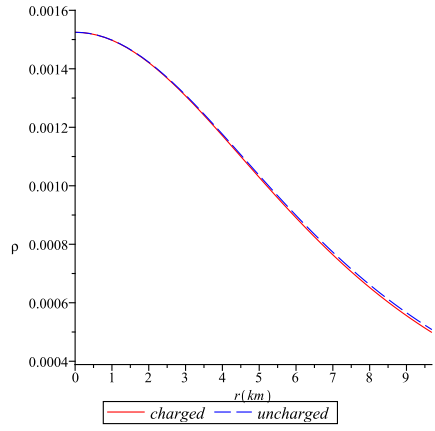}
       \caption{The matter density is plotted against r for the star PSR J1614-2230.}
    \label{fig:1}
\end{figure}

\begin{figure}[htbp]
    \centering
        \includegraphics[scale=.3]{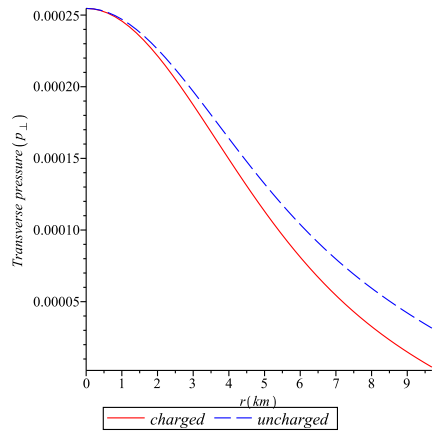}
       \caption{The transverse pressure $p_t$ is plotted against r for the star PSR J1614-2230.}
    \label{fig:3}
\end{figure}

\section{Physical Analysis}
\label{sec:4}
\noindent To be a physically acceptable model matter density $(\rho)$, radial pressure ($p_r$), transverse pressure ($p_\bot$) all should be non-negative inside the stellar interior. 
It is clear from equations (\ref{E2}) and (\ref{pr2}) it is clear that $pr$ is positive throughout the distribution. The profile of 
$\rho$ and $p_\bot$ are shown in fig. 1 and fig. 2 respectively. From the figure it is clear that all are positive inside the 
stellar interior.\\
The profile of $\frac{d \rho}{dr},~\frac{d p_r}{dr}$ and $\frac{dp_{\perp}}{dr}$ are shown in fig. 3, it is clearly indicates
that $\rho$, $p_{r}$ and $p_{\perp}$ are descreasing in radially outward direction.
According to \cite{bondi} for an anisotropic fluid sphere the trace of the energy
tensor should be positive. To check this condition for our model we plot $\rho-p_r-2p_{\perp}$ against r in Fig. 4. From the 
figure it is clear that our proposed model of compact star satisfies Bondi's conditions.

\begin{figure}[htbp]
    \centering
        \includegraphics[scale=.3]{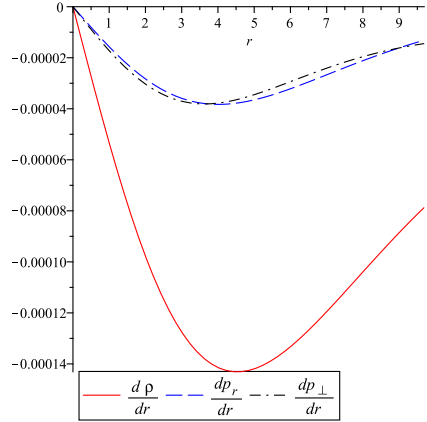}
       \caption{$\frac{d\rho}{dr}$, $\frac{dp_r}{dr}$ and $\frac{dp_{\perp}}{dr}$ are plotted against r for the star PSR J1614-2230. }
    \label{fig:4}
\end{figure}

\begin{figure}[htbp]
    \centering
        \includegraphics[scale=.3]{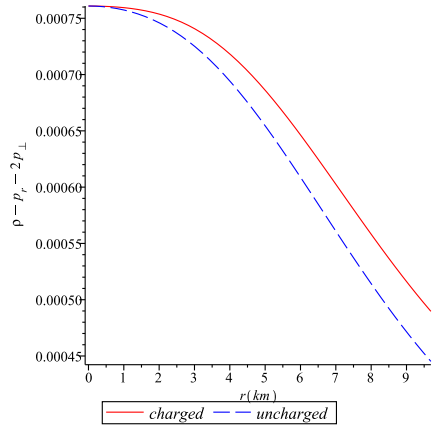}
       \caption{$\rho-p_r-2p_t$ is plotted against r for the star PSR J1614-2230.}
    \label{fig:5}
\end{figure}

\noindent For a physically acceptable model of anisotropic fluid sphere the radial and transverse velocity of sound should be 
less than 1 which is known as causality conditions.\\
Where the radial velocity $(v_{sr}^{2})$ and transverse velocity $(v_{st}^{2})$ of sound can be obtained as
\begin{equation}\label{dprdrho}
	\frac{dp_{r}}{d\rho}=\frac{bp_{0}(3-ar^{2})}{5b+\alpha+a(b-\alpha)r^{2}}.
\end{equation}
\begin{equation}\label{dppdrho}
	\frac{dp_{\perp}}{d\rho}=\frac{(1+ar^{2})^{3}\left[D_{1}+D_{2}r^{2}+D_{3}r^{4}+D_{4}r^{6}+D_{5}r^{8} \right]}{\left[-10ab-2a\alpha-2a^{2}(b-\alpha)r^{2} \right]\left[2+E_{1}r^{2}+E_{2}r^{4}+E_{3}r^{6}+E_{4}r^{8}+E_{5}r^{10}+E_{6}r^{12} \right]}.
\end{equation}
where,\\
$D_{1}=b^{2}p_{0}^{2}+4b^{2}p_{0}-24abp_{0}+3b^{2}-12\alpha a$,\\
 $D_{2}=-6ab^{2}p_{0}^{2}+32ab^{2}p_{0}-24a^{2}bp_{0}-4\alpha abp_{0}-2ab^{2}+16\alpha ab-8\alpha a^{2}$,\\
$D_{3}=5ab^{3}p_{0}^{2}-8ab^{3}p_{0}+2\alpha ab^{2}p_{0}-12a^{2}b^{2}p_{0}+24a^{3}bp_{0}+6\alpha a^{2}bp_{0}+7ab^{3}-12a^{2}b^{2}-8\alpha ab^{2}-8\alpha a^{2}b+24\alpha a^{3}+3\alpha^{2}a^{2}$,\\
$D_{4}=6a^{3}b^{b}p_{0}^{2}-6a^{2}b^{3}p_{0}^{2}+16a^{2}b^{3}p_{0}-40a^{3}b^{2}p_{0}-8\alpha a^{2}b^{2}p_{0}+24a^{4}bp_{0}+8\alpha a^{3}bp_{0}+6a^{2}b^{3}+8\alpha a^{2}b^{2}-6a^{3}b^{2}-32\alpha a^{3}b-2\alpha^{2}a^{2}b+24\alpha a^{4}+2\alpha^{2}a^{3}$,\\
$D_{5}=a^{3}b^{3}p_{0}^{2}-a^{4}b^{2}p_{0}^{2}+2\alpha a^{3}b^{2}p_{0}-2\alpha a^{4}bp_{0}-a^{3}b^{3}+a^{4}b^{2}+4\alpha a^{3}b^{2}+\alpha^{2}a^{3}b-8\alpha a^{4}b+4\alpha a^{5}-\alpha^{2}a^{4}$,\\
$E_{1}=12a-4b$,~~~~ $E_{2}=2b^{2}-20ab+30a^{2}$,~~~~ $E_{3}=8ab^{2}-40a^{2}b+40a^{3}$,~~~~ $E_{4}=12a^{2}b^{2}-40a^{3}b+30a^{4}$,\\
$E_{5}=8a^{3}b^{2}-20a^{4}b+12a^{5}$ ~~~~~and~~ $E_{6}=2a^{4}b^{2}-4a^{5}b+2a^{6}$.\\

Due to the complexity of the expression of $v_{st}^{2}$ we prove the causality conditions with the help of graphical 
representation. The graphs of $(v_{sr}^{2})$ and $(v_{st}^{2})$ have been plotted in fig. 5 and fig. 6 respectively. 
From the figure it is clear that $0 <v_{sr}^{2}\leq 1$ and $0<v_{st}^{2} \leq 1$ everywhere within the stellar configuration.
Moreover $\frac{dp_t}{d\rho} $ and$\frac{dp_r}{d\rho} $ are monotonic decreasing function of radius `r' for $0\leq r \leq r_b$ 
which implies that the velocity of sound is increasing with the increase of density.

\noindent A relativistic star will be stable if the
relativistic adiabatic index $\Gamma>\frac{4}{3}$. where $\Gamma$ is given by
\begin{equation}
\Gamma=\frac{\rho+p_r}{p_r}\frac{dp_r}{d\rho}
\end{equation}
To see the variation of the relativistic index we plot $\Gamma$ for our present of compact star which is plotted in fig. 7. 
The figure ensures that our model is stable.
\begin{figure}[htbp]
    \centering
        \includegraphics[scale=.3]{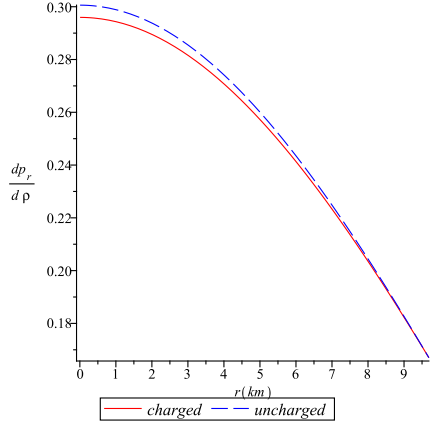}
       \caption{$v_{sr}^{2}=\frac{dp_r}{d\rho}$ is plotted against r for the star PSR J1614-2230.}
    \label{fig:6}
\end{figure}

\begin{figure}[htbp]
    \centering
        \includegraphics[scale=.3]{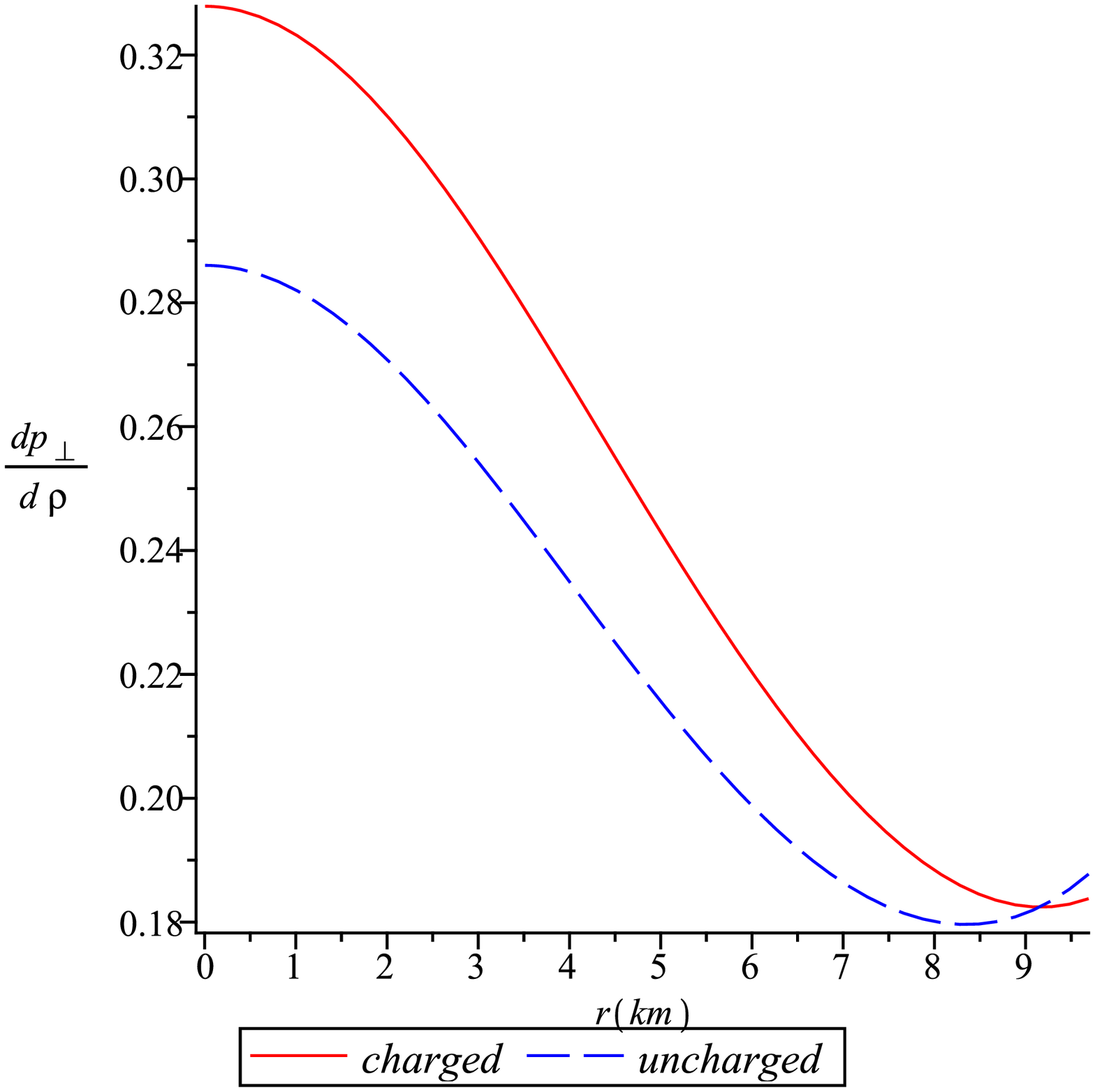}
       \caption{$v_{st}^{2}=\frac{dp_\perp}{d\rho}$ is plotted against r for the star PSR J1614-2230.}
    \label{fig:7}
\end{figure}

\begin{figure}[htbp]
    \centering
        \includegraphics[scale=.3]{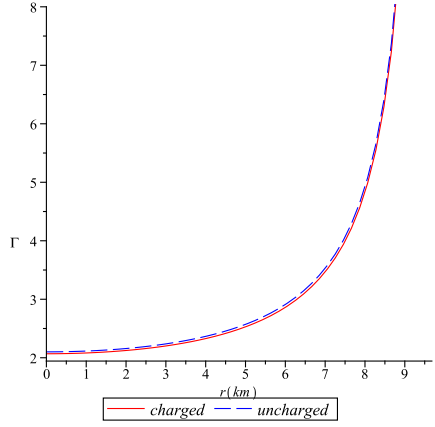}
       \caption{The adiabatic index $\Gamma$ is plotted against r for the star PSR J1614-2230.}
    \label{fig:8}
\end{figure}

\begin{table}
\small
\caption{The values of `a' and `b' obtained from the equation (\ref{a}) and (\ref{b})}
{\begin{tabular}{@{}crrrrrrrrrrr@{}} \hline
Compact Star &&$ M(M_\odot)$ &Mass(km)& Radius(km) & a($km^{-2}$) & b($km^{-2}$) & u & $z_s$\\ \hline
4U 1820-30 &&1.58&	2.33050&	 9.1&	  0.012076&	 0.012370 &   0.256099 & 0.431786 \\
PSR J1903+327 &&1.667&	2.45882&	 9.438&	  0.011226&	 0.011699&   0.260524& 0.444954   \\
4U 1608-52 &&1.74&	2.56650&	 9.31&	  0.011537&	 0.012722&   0.275671 & 0.492941  \\
Vela X-1&& 1.77&	2.61075&	 9.56&	  0.010942&	 0.011952 &  0.273091 & 0.484428   \\
PSR J1614-2230 && 1.97&	2.90575&	 9.69&	  0.01065&	 0.012775&   0.299871&  0.580629   \\
Cen X-3&&1.49&	2.19775&	 9.178&	  0.011871&	 0.011371&   0.239458 & 0.385309  \\ \hline
\end{tabular} \label{table1}}
\end{table}

\begin{table}
\small
\caption{The values of central density, surface density, central pressure and radial velocity of the sound at the origin for different compact stars are obtained.}
{\begin{tabular}{@{}crrrrrrrrrrr@{}} \hline
Compact Star && central density~$(\rho_0)$  &surface density& surface density & central pressure~$(p_0)$ & $\frac{dp_r}{d\rho}_{|r=0}$\\
  &&               $~gm.cm^{-3}$& (uncharged) & (charged)& $~dyne.cm^{-2}$& (charged)\\ \hline
4U 1820-30&& 1.994 $\times 10^{15}$ &   6.648 $\times 10^{14}$&      6.514 $\times 10^{14}$ &   2.989 $\times 10^{35}$ & 0.295227\\
PSR J1903+327&& 1.886 $\times 10^{15}$&   6.287 $\times 10^{14}$&      6.153 $\times 10^{14}$&    2.827 $\times 10^{35}$& 0.294958\\
4U 1608-52&& 2.051 $\times 10^{15}$&   6.837 $\times 10^{14}$ &     6.703 $\times 10^{14}$ &   3.074 $\times 10^{35}$& 0.295357\\
Vela X-1 &&1.927 $\times 10^{15}$ &  6.423 $\times 10^{14}$&      6.289 $\times 10^{14}$&    2.888 $\times 10^{35}$& 0.295063\\
PSR J1614-2230&& 2.059 $\times 10^{15}$&   6.865 $\times 10^{14}$&      6.731 $\times 10^{14}$&    3.087 $\times 10^{35}$ & 0.295376\\
Cen X-3 &&1.833 $\times 10^{15}$ &  6.111 $\times 10^{14}$ &     5.977 $\times 10^{14}$ &   2.748 $\times 10^{35}$ & 0.294815\\  \hline
\end{tabular} \label{ta1}}
\end{table}

\noindent For an anisotropic fluid sphere all the energy conditions namely Weak Energy Condition (WEC), 
Null Energy Condition (NEC), Strong Energy Condition (SEC) and Dominant Energy Condition (DEC) are satisfied if and only if the 
following inequalities hold simultaneously in every point inside the fluid sphere.
\begin{figure*}[thbp]
\begin{center}
\begin{tabular}{rl}
\includegraphics[width=4.6cm]{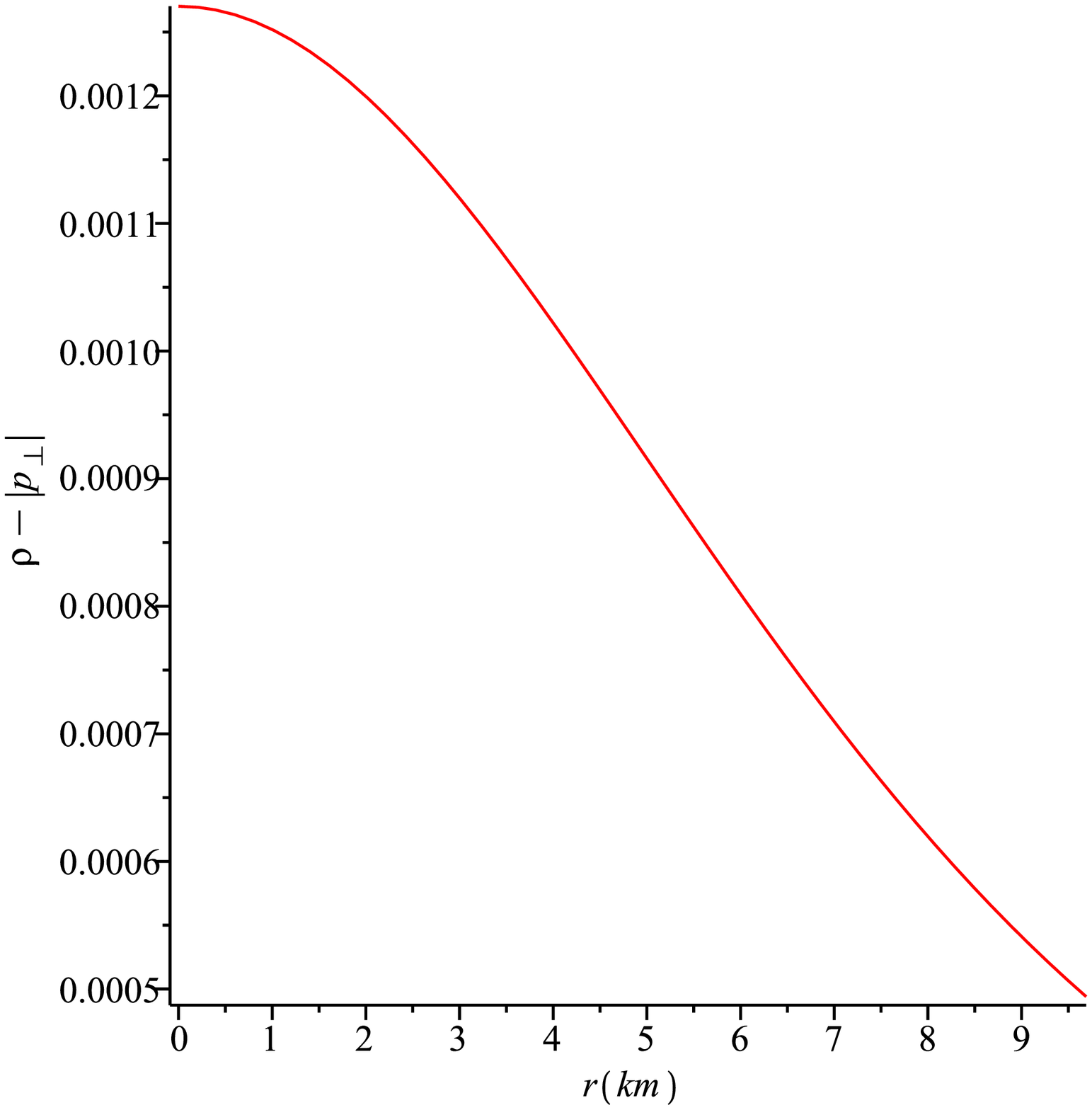}&
\includegraphics[width=4.6cm]{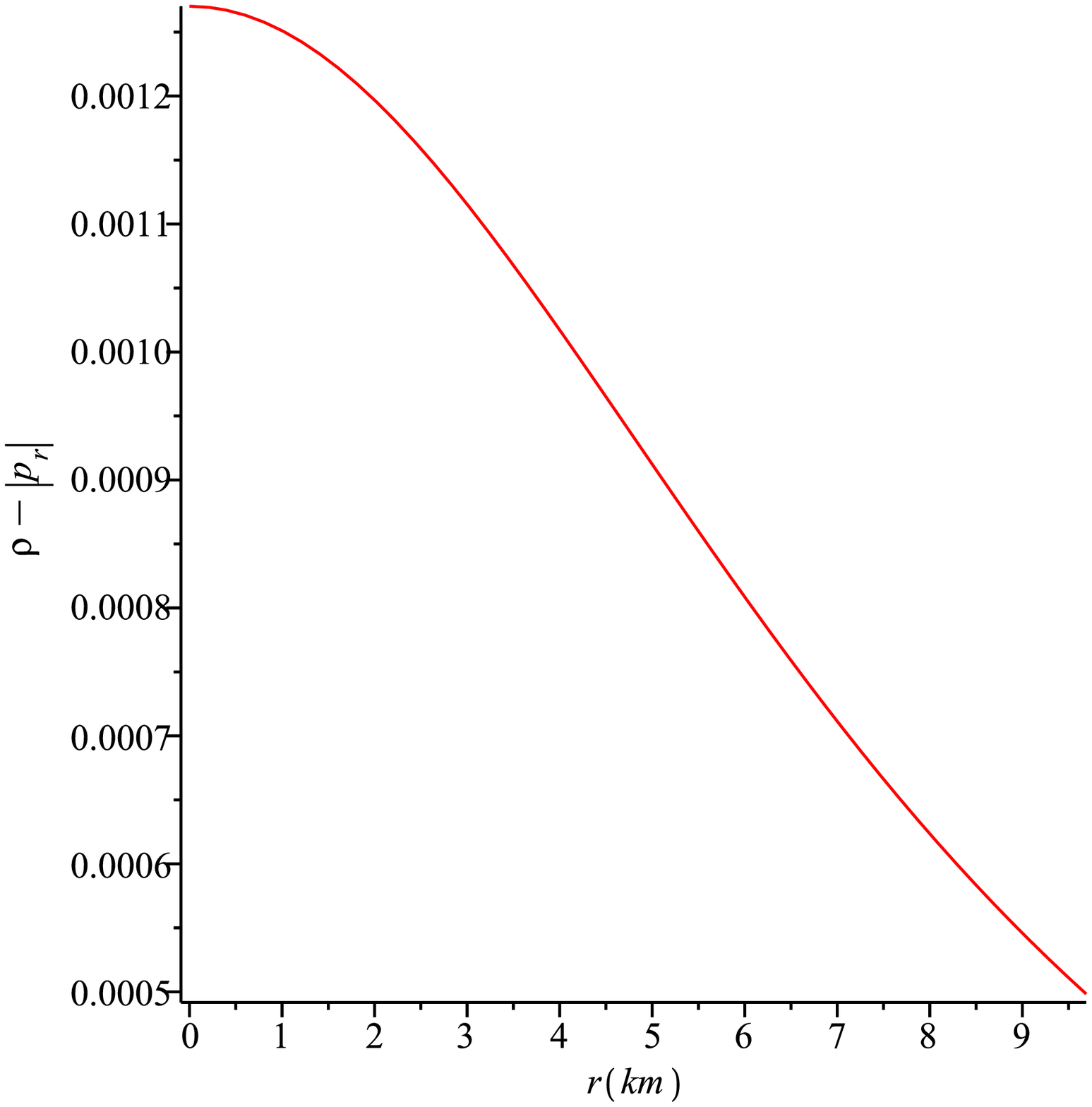}
\includegraphics[width=4.6cm]{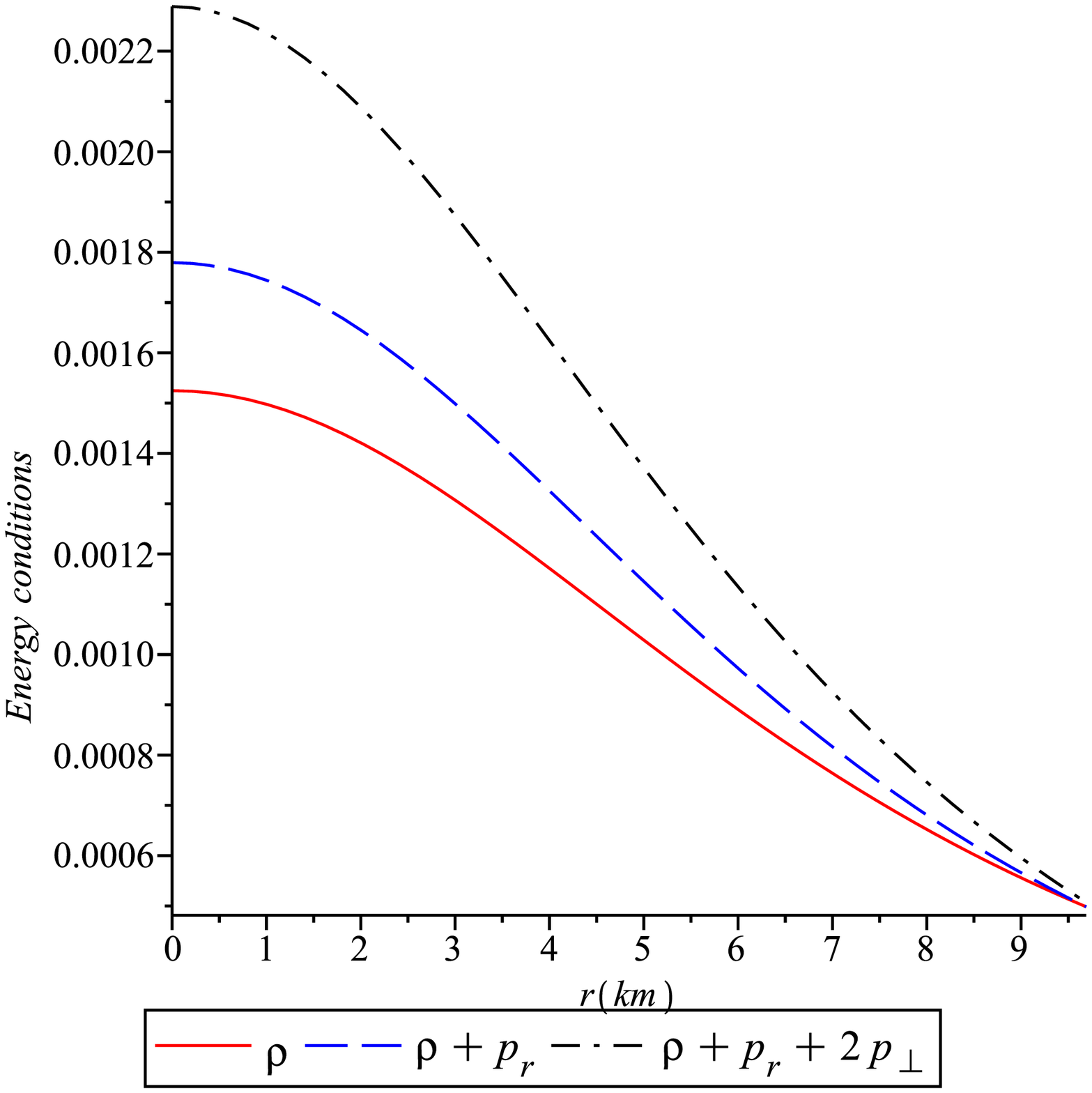}\\
\end{tabular}
\end{center}
\caption{The left and middle figures show the dominant energy conditions where as the right figure shows the weak null and strong energy conditions are satisfied by our model for the star PSR J1614-2230. }
\end{figure*}

\begin{equation}\label{ec1}
(i)NEC:\rho+p_r\geq 0
\end{equation}
\begin{equation}\label{ec2}
(ii)WEC:p_r+\rho\geq 0,~~~\rho>0
\end{equation}
\begin{equation}\label{ec3}
(iii)SEC:\rho+p_r\geq 0~~~~\rho+p_r+2p_{\perp}\geq 0
\end{equation}
\begin{equation}\label{ec4}
(iv)DEC:\rho >\left|p_r\right| ~~~, \rho >\left|p_{\perp}\right|
\end{equation}

\noindent Due to the complexity of the expression of $p_{\perp}$ we will prove the inequality (\ref{ec1})-(\ref{ec4}) 
with the help of graphical representation. The profiles of the L.H.S of the above inequalities are depicted in fig.~8 for the 
compact star PSR J1614-2230. The figure shows that all the energy conditions are satisfied by our model of compact star.

\begin{figure}[htbp]
    \centering
        \includegraphics[scale=.3]{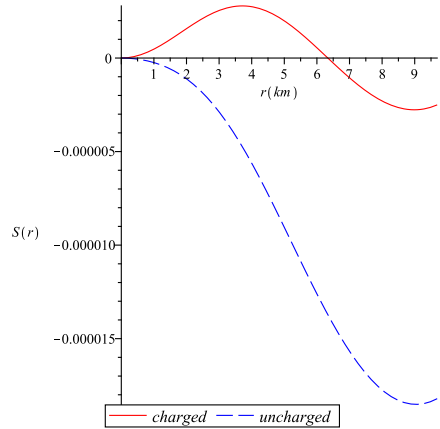}
       \caption{Variation of anisotropy is shown against r for the star PSR J1614-2230.}
    \label{fig:10}
\end{figure}

\noindent The ratio of mass to the radius of a compact star can not be arbitrarily large. \cite{buchdahl59} showed that  for a 
(3+1)-dimensional fluid sphere $\frac{2M}{r_b}<\frac{8}{9}$. To see the maximum ratio of mass to the radius  for our model we 
calculate the compactness of the star given by
\begin{equation}
u(r)=\frac{m(r)}{r}=\frac{br^{2}}{2(1+ar^{2})},
\end{equation}
and the corresponding surface redshift $z_s$ is obtained by,
\[1+z_s(r_b)=\left[1-2u(r_b)\right]^{-1/2}\].
Therefore $z_s$ can be obtained as,
\begin{equation}
z_s(r_b)=\left[\frac{1+(a-b)r_b^{2}}{1+ar_b^{2}}\right]^{-\frac{1}{2}}-1.
\end{equation}
The surface redshift of different compact stars are given in table \ref{table1}.

\begin{figure}[htbp]
    \centering
        \includegraphics[scale=.3]{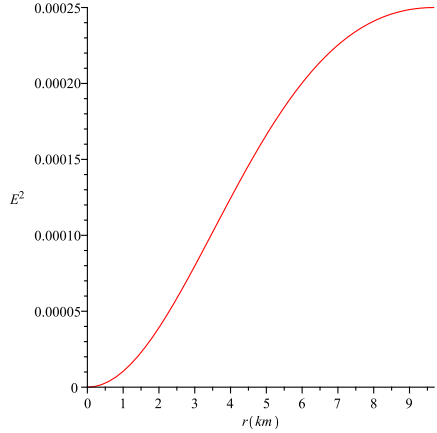}
       \caption{The variation of electric field is shown against r for the star PSR J1614-2230.}
    \label{fig:11}
\end{figure}

\section{Conclusion}
\label{sec:6}
We have obtained a new class of solution for charged compact stars having \cite{mat80} mass function. The electric field 
intensity is increasing in radially outward direction and the adiabatic index $\Gamma>\frac{4}{3}$. The physical requirements are
checked for the star PSR J1614-2230 and model satisfies all the physical conditions. Some salient features of the model are 
\begin{itemize}
	\item[(i)] In present model if $\alpha=0$, the model corresponds to \cite{ratanpal4} model.
	\item[(ii)] In present model if $\alpha=0$, $a=b=\frac{1}{R^{2}}$, where $R$ is geometric parameter then the model corresponds
                    to \cite{sharma13} model, which is stable for $\frac{1}{3}<p_{0}<0.3944$.
\end{itemize}

\section*{Acknowledgement}
\noindent BSR is thankful to IUCAA, Pune, for providing the facilities and hospitality where the part of this
work was done.

\end{document}